\documentstyle[12pt,twoside,epsfig]{article}
\begin{document}
%%%%%%%%%%%%%%%%%%%%%%%%%%%%%%%%%%%%%%%%%%%%%%%%%%%%%%%%%%%%%%

\begin{center}
{\Large \bf   The polaron confined in one dimension }\\[0.5cm]

                            G.~Ganbold
\footnote{E-mail: ganbold@thsun1.jinr.ru;
{\it Permanent address:} Institute of Physics and Technology,
           Mongolian Academy of Sciences, 210651 Ulaanbaatar, Mongolia}
\vskip 5mm

{\footnotesize Bogoliubov Laboratory of Theoretical Physics,
Joint Institute for Nuclear Research, \\ 141980 Dubna, Russia}
\end{center}

\vskip 2mm

\begin{abstract}
 The ground-state energy, the effective mass and the number of
virtual phonons of the optical large polaron confined strictly in
one dimension have been estimated by using the generalized Gaussian
approximation. The leading-order terms take care of all Gaussian
fluctuations in the system and improve the conventional variational
estimates at finite coupling. Particularly, the lowest upper bound to
the polaron ground-state energy has been obtained. The non-Gaussian
contributions systematically correct the leading-order approximations.
We have obtained exact analytical solutions in the weak- and
strong-coupling limit and reasonable numerical data for intermediate
coupling. Our result for the number of excited phonons limits the
validity region of the few-phonon approximation methods.  \\
\end{abstract}

\vskip 2mm

\noindent
PACS number(s): 31.15.Kb; 63.20.Kr; 71.38.+i; 73.20.Dx \\[2mm]

%%%%%%%%%%%%%%%%%%%%%%%%%%%%%%%%%%%%%%%%%%%%%%%%%%%%%%%%%%%%%%%%%%%%
\section{ Introduction }
%===================================================================

 A conducting electron in a polar crystal polarizes the surrounding
lattice and, by interacting  with the quanta of lattice vibration,
phonons, forms a quasi-particle, the {\sl polaron}.

The polaron problem traditionally has been investigated in three
dimensions \cite{froh54,feyn55}, although the electron motions may be
localized in reduced dimensionality \cite{lork90}. Many theoretical
investigations have been made to study the polaronic properties in
lower dimensions (see, e.g. \cite{gros76,lesc89}). A considerable
research attention has been devoted to the study of the polaron
confined in one dimension \cite{dega86,peet91,quin94,ganb00}. In
particular, recent studies on the electron-phonon interaction in
quasi one-dimensional systems have explored a polaronic effect
pronounced stronger than in two-dimensional structures \cite{hai93}.
There are two distinct models employed for the one-dimensional
polaron: the small and the large polaron. Small polarons are more
strictly bounded and observed in linear conjugative organic polymer
conductors \cite{suyu83} while the latter model describes electrons
confined in one-direction in polar semiconductors \cite{sern87}. The
phonons have been considered in the optical and acoustic modes,
depending on the specific dispersion rule \cite{dega86}. Regardless
of the  extremely confined structure of the system, polar optical
phonons are treated in the spirit of bulk theory.

The resulting interaction of an electron confined in one dimension
with the phonons of longitudinal optical lattice vibrations can be
modelled by the following Hamiltonian \cite{dega86,peet91,quin94}
\begin{equation}
H=\frac{p^2}{2m} + \hbar\omega_{0}\sum_{k} a^{\dagger}_{k} a_{k}
+ g\sum_{k}\Big( a_{k} e^{i k r} + a^{\dagger}_{k} e^{-i k r}\Big)\,,
\qquad g = \hbar\omega_{0} \left( \hbar \, \alpha/ m \,\omega_{0}\,L
\right)^{1/2} \,,
\label{hamilt}
\end{equation}
where $m$, ${p}$ and ${r}$ denote the electron bare mass, the
momentum and position of the electron, $L$ is the length of lattice
crystal, whereas $k$, $a_{k}$ and $a^{\dagger}_{k}$ are wave vector,
annihilation and creation operators of a phonon. The optical polaron
model imposes the phonon frequency $\omega_{0}$ independent on $k$.
Further we will use appropriate units, such that,
$m=\omega_{0}=\hbar=1$. The electron-phonon coupling factor $g$
behaves $g\propto \sqrt{\alpha/\Omega}/ |k|^{(d-1)/2}$ in $d$
dimensions \cite{peet86,ganb94}, where $\Omega$ is the volume of the
system. Therefore, $g$ does not depend on $k$ in one dimension
\cite{dega86,quin94}. The Fr\"ohlich dimensionless coupling constant
$\alpha$ takes values ranging from 0.02 to 4 for most actual crystals.
In this paper the coupling constant $\alpha$ is identical to
$2\pi\alpha_{op}$ in \cite{dega86}, $\alpha'$ in \cite{peet91} and
$\alpha$ in \cite{quin94}.

Many approaches have been developed to investigate the quasi-particle
properties of the system (\ref{hamilt}) in the ground state
\cite{dega86,peet91}. Strict results for these quantities are
available only in the weak- ($\alpha\to 0$) and strong-coupling
($\alpha\to\infty$) limit (see, e.g. \cite{gros76,quin94}).

In the present paper we have made a systematic investigations of the
large optical polaron confined strictly in one dimension within a
path-integral (PI) method. We have obtained exact analytical solutions
of the self-energy, the effective mass and the number of virtual
phonons in the weak- and strong-coupling limit as well as reasonable
numerical data improving known results \cite{dega86,peet91,quin94} at
finite $\alpha$. Particularly, our result for the number of excited
phonons limits the validity region of the few-phonon approximations
\cite{quin94} and shows that as $\alpha$ grows, the efficiency of the
general Gaussian  approximation drops faster than in higher dimensions.

%%%%%%%%%%%%%%%%%%%%%%%%%%%%%%%%%%%%%%%%%%%%%%%%%%%%%%%%%%%%%%%%%%%%
\section{ Polaron Ground-state Properties }
%===================================================================

{\bf 1.} The phonon variables in (\ref{hamilt}) may be exactly
eliminated by using the Feynman PI technique \cite{feyn55}. The
partition function for a resting polaron reads
\begin{eqnarray}
\label{PIZ}
&& Z_{\beta}(\alpha) = e^{-\beta F(\alpha,\beta)} = \left\langle
e^{\alpha W[r]} \right\rangle_{0}\,, \qquad \left\langle (\ldots)
\right\rangle_{0} = \int\limits_{r(0)=0}^{r(\beta)=0}\!\!\!\!\!\!
\delta r \, e^{-S_{0}[r]} (\ldots) \,,                         \\
&& S_{0}[r] = {1\over 2} \int\limits_{0}^{\beta}\!dt ds \,
r(t)\, D_0^{-1}(t,s) \, r(s) \,, \qquad
D_0^{-1}(t,s) = - \delta(t-s) \, {d^2 \over dt^2}  \,,\nonumber
\end{eqnarray}
where the inverse temperature $\beta=1/{\it k}_B T$ is infinite in the
ground state: $\beta\to\infty$. Therefore, $E(\alpha)=F(\alpha,\infty)$
represents the polaron {\sl ground-state energy} (GSE). The polaron
retarded self-interaction is given by
\begin{eqnarray}
\label{retarded}
W[r] = {1\over\sqrt{2}} \int\!\!\!\!\int\limits_{0}^{\beta}\!\! dt ds
\, e^{-|t-s|} \int\limits_{-\infty}^{\infty}\!\! {dk\over 2\pi}
\, e^{ikR(t,s)} = \int\!\! d\Omega_{tsk} \, e^{ikR(t,s)} \,,
\qquad R(t,s) = r(t)-r(s) \,.
\end{eqnarray}

{\bf 2.} For a slow polaron the partition function projected at small
fixed momentum $p$ is
\begin{equation}
Z_{\beta}(\alpha,p)=e^{-\beta E(\alpha,p)}={1\over\sqrt{2\pi\beta}}
\int\!\! dx\, e^{-ipx} \!\!\!\int\limits_{\rho(0)=0}^{\rho(\beta)=x}
\!\!\!\!\!\! \delta \rho \, e^{-S_{0}[\rho] + \alpha W[\rho]} \,.
\label{partfun}
\end{equation}
Since the polaron action is translationally invariant, the energy
$E(\alpha,p)$ is a continuous function of $p$. By expanding it around
small momentum we define the {\sl effective mass} (EM) of the polaron
$m^*(\alpha)$ as follows:
\begin{equation}
\label{expa}
E(\alpha,p) = E(\alpha) + {p^2 \over 2 m^*(\alpha)} + O(p^4) \,.
\end{equation}
With the new integration variable $r(t)=\rho(t)-q(t)$, where
$q(t)=xt/\beta=yt$, one goes to conventional closed-end paths
$r(t)$ and rewrites (\ref{partfun}) as follows:
\begin{equation}
e^{-\beta E(\alpha,\, p)} = \sqrt{\beta\over 2\pi} \int\!\! dy \,
e^{-i\beta py-\beta y^2/2} \, e^{-\beta {\cal F}(\alpha,\, y^2)} \,,
\qquad  e^{-\beta {\cal F}(\alpha,\, y^2)}
= \left\langle  e^{\alpha W[r+q]} \right\rangle_{0}  \,.
\label{purpose}
\end{equation}
For $\beta\to\infty$ the integral over $y$ in (\ref{purpose}) may be
taken by the saddle point method. The saddle point $y_0=\sqrt{\xi}$
satisfies the equation
\begin{equation}
ip + \sqrt{\xi} \,\left[ 1 + 2\, {\cal F}_{\xi}(\alpha,\xi) \right]
= 0 \,, \qquad {\cal F}_\xi(\alpha,\xi)
= {d\over d\xi} {\cal F}(\alpha,\xi)   \,.
\label{saddle1}
\end{equation}
Then,
\begin{equation}
E(\alpha,p) = i p \sqrt{\xi} + {\xi\over 2} + {\cal F}(\alpha,\xi)
= {\cal F}(\alpha) + {p^2\over 2\left( 1 + 2\,{\cal F}_\xi(\alpha)
\right)} + O(p^4) \,,
\label{extra}
\end{equation}
where ${\cal F}(\alpha)={\cal F}(\alpha,0)$ and
${\cal F}_\xi(\alpha)={\cal F}_\xi(\alpha,0)$. Comparing (\ref{expa})
with (\ref{extra}) we find
\begin{equation}
\label{enermass}
E(\alpha) = {\cal F}(\alpha) \,, \qquad
m^{*}(\alpha) = 1 + 2 {\cal F}_\xi(\alpha) \,.
\end{equation}
To evaluate ${\cal F}_\xi(\alpha)$ we use the expansion
\begin{eqnarray*}
W[r+q] &=& W[r]+iy\, V_1[r]-{y^2\over 2}\, V_2[r]+O(y^3) \,,\\
V_1[r] &=& \int\! d\Omega_{tsk}\, e^{ikR(t,s)}\, k\, (t-s) \,,
\qquad V_2[r] = \int\! d\Omega_{tsk}\, e^{ikR(t,s)}\, k^2\,(t-s)^2
\,. \nonumber
\end{eqnarray*}
Neglecting terms $\sim O(y^3)$ we obtain
\begin{eqnarray}
e^{-\beta {\cal F}(\alpha,y^2)} = Z_{\beta}(\alpha)\, \exp\left\{-y^2
{Y_{\beta}(\alpha)\over Z_{\beta}(\alpha)} \right\} \,, \qquad
Y_{\beta}(\alpha)={1\over 2} \left\langle \, e^{\alpha W[r]} \,
\left( V_1^2 [r] + V_2 [r] \right) \right\rangle_0  \,.
\label{PIY}
\end{eqnarray}
Hence,
\begin{eqnarray}
{\cal F} (\alpha) = - \lim\limits_{\beta\to\infty} {1\over\beta}
\ln Z_{\beta}(\alpha) \,, \qquad
{\cal F}_\xi(\alpha) = \lim\limits_{\beta\to\infty}
{Y_{\beta}(\alpha)\over Z_{\beta}(\alpha)} \,.
\label{massint}
\end{eqnarray}

{\bf 3.} The {\sl average number of virtual phonons} (ANVP) excited
by the electron-lattice interaction is given by
\begin{equation}
\label{numb}
N(\alpha) = \langle 0| \sum\limits_{k} a^{\dagger}_{k} a_{k}
|0\rangle = \left.\left\langle {\partial H \over\partial\omega_0}
\right\rangle_{0} \right\vert_{\omega_0=1} = -{1\over\beta}
\left. {\partial\over\partial\omega_0} \ln \mbox{Tr} e^{-\beta H}
\right\vert_{\omega_0=1}  \,.
\end{equation}
The differentiation in (\ref{numb}) should not affect the interaction
term in $H$. We obtain
\begin{eqnarray}
N(\alpha) = {\alpha\over \sqrt{2}\,\pi} \!\int\limits_{0}^{\infty}
\!\! d\tau \,\tau\, e^{-\tau} \!\!\int\limits_{-\infty}^{\infty} \!\!
dk \, {X_{\beta}(\alpha,k,\tau) \over Z_{\beta}(\alpha)} \,, \quad
X_{\beta}(\alpha,k,\tau) = \left\langle e^{ikR(t,s)+\alpha W[r]}
\right\rangle_{0} \,,
\label{number}
\end{eqnarray}
where $\tau=t-s$. Note, the following relation between the ANVP and
GSE takes place
\cite{smon87}
\begin{equation}
\label{number2}
N(\alpha) = \left( 1 - {3\over 2}\, \alpha \,
{\partial\over\partial\alpha} \right) E(\alpha) \,.
\end{equation}

%%%%%%%%%%%%%%%%%%%%%%%%%%%%%%%%%%%%%%%%%%%%%%%%%%%%%%%%%%%%%%%%%%%%
\section{Generalized Gaussian Approximation}
%===================================================================

 For finite $\alpha$, the nonlocality and $\delta$-singularity arising
in (\ref{retarded}) prevent any explicit evaluation of
$Z_{\beta}(\alpha)$, $Y_{\beta}(\alpha)$ and
$X_{\beta}(\alpha,k,\tau)$. Below we represent a nonvariational
approximation method, which isolates completely the Gaussian
contributions from these PIs. The remaining non-Gaussian
corrections turn to be relatively small (at least for the energy) and
may be systematically estimated to improve the leading-order
approximations.

{\bf i.}  First, we demonstrate the basic idea of our method by
evaluating $Z_{\beta}(\alpha)$. Note, the initial representation
(\ref{PIZ}) is optimal only for $\alpha\to 0$. To describe the system
more efficiently at finite $\alpha$ we introduce a functional
integration weighted with the most general Gaussian measure
$d\sigma[r]$ as follows
\begin{equation}
\langle (\ldots) \rangle = \int\limits_{r(0)=0}^{r(\beta)=0}
\!\!\!\! d\sigma[r] \, (\ldots) \,, \qquad
d\sigma[r] = \delta r \, e^{-{1\over 2} (r, \, D^{-1} \, r)}
\,, \qquad \langle 1 \rangle = 1 \,.
\label{measur2}
\end{equation}
We assume that all Gaussian configurations of the system is totally
concentrated in $d\sigma[r]$. This imposes specific requirements on
$D^{-1}(t,s)$ (and on its Green function $D(t,s)$) that will be
discussed later. Obviously, the following relations take place
\begin{eqnarray*}
\langle r(t)\, r(s) \rangle = D(t,s) \,, \quad
\langle e^{i k R(t,s)} \rangle = e^{-k^2 F(|t-s|)} \,, \quad
F(t,s) = {D(t,t)+D(s,s)\over 2} - D(t,s) \,.
\end{eqnarray*}
Then, we rewrite (\ref{PIZ}) in the new representation as follows
\begin{equation}
Z_{\beta}(\alpha)= \exp\left\{ {1\over 2}\mbox{Tr}\ln {D\over D_{0}}
\right\} \left\langle \exp\left\{
{1\over 2}\!\int\!\!\!\!\int\limits_{0}^{\beta}\! dt ds \, r(t) \,
D^{-1}(t,s)\, r(s) - S_0[r] +\alpha\, W[r] \right\}\right\rangle \,.
\label{ident}
\end{equation}

 Let us introduce the conception of the {\it normal-ordered form} of
functionals with respect to $d\sigma[r]$. In particular, we use the
following normal forms:
\begin{eqnarray}
\label{normform}
:\! r(t)~r(s) \!: = r(t)~r(s) - D(t,s) \,, \quad
:\!e^{ikR(t,s)}\!: = \! e^{ikR(t,s)}\, e^{k^2 F(t,s)}
\end{eqnarray}
so that
\begin{eqnarray}
\langle :\! r(t)~r(s)\!: \rangle = 0\,, \qquad
\langle :\!e_2^{ikR(t,s)}\!: \rangle = 0 \,, \qquad
e^{z}_2 = e^{z}-1-z-{z^2\over 2} \,.
\end{eqnarray}
Functional $W[r]$ may be decomposed as follows:
\begin{eqnarray}
\label{intWD}
&& W[r] = \int\!\! d\Lambda_{tsk} \, :\!e^{ikR(t,s)}\!: \,\,
= W[0]+iW_{1}[r]-{1\over 2}W_{2}[r]+W_{int}[r]  \,,       \nonumber\\
&& W_{1}[r] = \int\!\! d\Lambda_{tsk} \, k :\! R(t,s) \!:  \,,\qquad
\qquad W_{2}[r]=\int\!\! d\Lambda_{tsk}\,k^2 :\! R^2(t,s) \!: \,, \\
&& W_{int}[r] =\int\!\! d\Lambda_{tsk}~:\! e_2^{ikR(t,s)}\!: \,,\qquad
\qquad d\Lambda_{tsk}=d\Omega_{tsk} \, e^{-k^2 F(t,s)} \,. \nonumber
\end{eqnarray}

 Since all quadratic configurations of the polaron action in the new
representation is totally included in $d\sigma[r]$, any extra quadratic
parts should be eliminated as follows
\begin{eqnarray}
{1\over 2} \int\!\!\!\!\int\limits_{0}^{\beta}\! dt ds \, :r(t)
\left[ D^{-1}(t,s)-D_{0}^{-1}(t,s)\right] r(s):
+ \, \alpha \,  W_2[r] = 0\,, \qquad \forall r \,.
\label{constraint}
\end{eqnarray}
This requirement leads to the following constraint equations for
function $F(t)$:
\begin{eqnarray}
F(t) &=& {1\over\pi}\int\limits_{0}^{\infty}\!\!dk \,
[1-\cos(kt)]\, \widetilde{D}(k) \,,               \nonumber\\
\widetilde{D}(k) &=& \left( k^2+{\alpha\over\sqrt{2\pi}}
\!\int \limits_{0}^{\infty}\!\!dt \, e^{-t}\,
{ 1-\cos(kt) \over F^{3/2}(t) } \right)^{-1}     \,,
\label{consteq}
\end{eqnarray}
where $\widetilde{D}(k)$ is the Fourier transform of $D(t)$. Note,
$\widetilde{D}_0(k)=1/k^2$. Taking into account (\ref{constraint})
we rewrite (\ref{ident})
\begin{eqnarray}
Z_{\beta}(\alpha) &=&  e^{-\beta E_o(\alpha)} \cdot
J_\beta(\alpha)\,, \qquad \qquad J_\beta(\alpha)
= \langle e^{\alpha \, W_{int}[r]} \rangle \,,       \nonumber\\
E_0(\alpha) &=& -\frac{1}{2\pi}\!\int\limits_{0}^{\infty}
\!\! dk \left[\,\ln\left(k^2{\widetilde D}(k)\right)
- k^2 {\widetilde D}(k) + 1 \right] + \frac{\alpha}{\sqrt{2\pi}}\!
\int\limits_{0}^{\infty} \!\!dt\, {\exp(-t)\over F^{1/2}(t)}\,,
\label{gga}
\end{eqnarray}
where $E_0(\alpha)$ is {\sl the Gaussian leading-order contribution}
to the GSE of the polaron confined in one dimension. The {\sl
non-Gaussian corrections} associated with $J_\beta(\alpha)$ can be
estimated systematically. Equations (\ref{ident})-(\ref{gga}) serve
as the basis of the Generalized Gaussian Approximation (GGA) method
applied to $Z_{\beta}(\alpha)$.

Note, $E_0(\alpha)$ represents a {\sl upper bound} to the true GSE of
the one-dimensional polaron. Indeed, $\langle W_{int}[r]\rangle = 0$.
Then, by using the Jensen-Peierls inequality one obtains
\begin{equation}
J_\beta(\alpha)\ge e^{\alpha\langle W_{int}[r]\rangle} = 1
\qquad \mbox{so,} \qquad  E(\alpha) \le  E_0(\alpha) \,.
\end{equation}

{\bf ii.} Now we evaluate $Y_{\beta}(\alpha)$ within the GGA method.
Remember, transformations (\ref{ident})-(\ref{gga}) affect only the
exponent in (\ref{PIY}), but not any prefactor of the exponential.
Therefore, despite the factor $\left( V_1^2[r]+V_2[r] \right)/2$, all
steps of the GGA method may be directly applied to $Y_{\beta}(\alpha)$.
Constraint equations (\ref{consteq}) remain without any change.
We obtain
\begin{eqnarray}
Y_{\beta}(\alpha) = e^{-\beta\,E_0(\alpha)} \cdot {1\over 2}
\left\langle e^{\alpha W_{int}[r]} \, \left( V_1^2[r]+V_2[r] \right)
\right\rangle
= e^{-\beta\,E_0(\alpha)} \left\{ {\cal F}_\xi^0(\alpha)
+ {1\over 2}{\cal J}_{\beta}(\alpha) \right\} \,,
\label{Ybeta}
\end{eqnarray}
where
\begin{eqnarray}
\label{calJbeta}
{\cal F}_\xi^0(\alpha) &=& {1\over 2} \int\!\! d\Lambda_{tsk}
\, k^2 \,(t-s)^2 \,, \qquad \qquad {\cal J}_{\beta}(\alpha)
= \left\langle e^{\alpha W_{int}[r]} \cdot Q[r] \right\rangle \,,  \\
Q[r] &=& \int\!\! d\Lambda_{tsk} \, k^2 (t-s)^2 \, \left( \,
e^{ikR(t,s)} - 1 \right)+\left[ \int\!\! d\Lambda_{tsk} \,
k\,(t-s) \left( \, e^{ikR(t,s)}-1 \right) \right]^2  \,. \nonumber
\end{eqnarray}
Substituting (\ref{Ybeta}) into (\ref{massint}) and (\ref{enermass})
we obtain the leading-order (Gaussian) approximation to the polaron
EM as follows
\begin{equation}
\label{gaussmass}
m^*_0(\alpha) =1 + 2 {\cal F}_{\xi}^{0}(\alpha)
= 1 + \frac{\alpha}{2\sqrt{2\pi}}\! \int\limits_{0}^{\infty} \!\!dt
\, t^2 \, {\exp(-t) \over F^{3/2}(t)} \,.
\end{equation}
Higher-order (non-Gaussian) corrections to the EM are given by
\begin{eqnarray}
\Delta m^*(\alpha)=\lim\limits_{\beta\to\infty} \,
{{\cal J}_{\beta}(\alpha) \over J_{\beta}(\alpha)}
- 2\,{\cal F}_\xi^0(\alpha)\cdot \left[ J_{\beta}(\alpha)-1 \right]\,.
\label{nongaussmass}
\end{eqnarray}

{\bf iii.} Note, the additional linear term $ikR(t,s)$ standing in the
exponent in (\ref{number}) affects neither the normal ordering, nor
the elimination of the quadratic path configurations. Hence, the
constraint equations (\ref{consteq}) do not change for the ratio
$X_{\beta}(\alpha,k,t)/Z_{\beta}(\alpha)$. So, we write
\begin{equation}
\label{PIXgauss}
{ X_{\beta}(\alpha,k,\tau) \over Z_{\beta}(\alpha)}
= e^{-k^2 F(\tau)} + \mbox{non-Gaussian~part}   \,.
\end{equation}

  Substituting (\ref{PIXgauss}) into (\ref{number}) and neglecting
the non-Gaussian part we obtain the leading-order approximation to
the ANVP as follows
\begin{equation}
\label{number0}
N_0(\alpha) = {\alpha\over \sqrt{2\pi}} \int\limits_{0}^{\infty}
\!\! dt \, e^{-t} \, {t \over F^{1/2}(t)} \,.
\end{equation}

%%%%%%%%%%%%%%%%%%%%%%%%%%%%%%%%%%%%%%%%%%%%%%%%%%%%%%%%%%%%%%%%%%%%
\section{Non-Gaussian Corrections}
%===================================================================

 To improve the obtained Gaussian approximations we use the following
expansions
\begin{eqnarray}
\label{expan}
J_\beta(\alpha) = \sum\limits_{n=2}^{\infty}
{\alpha^n \over n!} \, \langle W^n_{int}[r] \rangle  \,, \qquad
{\cal J}_\beta(\alpha) = \sum\limits_{n=0}^{\infty}
{\alpha^n \over n!}\,\langle W^n_{int}[r]\, Q[r] \rangle \,.
\end{eqnarray}
These are not plain perturbation expansions over $\alpha$ because each
of terms in angle brackets contains $\alpha$ in a complicated way so
that these series converge fast even for large $\alpha$. Taking into
account higher-order non-Gaussian contributions, we obtain corrected
values
\begin{eqnarray}
E_n(\alpha) &=& E_0(\alpha) + \Delta E_2(\alpha) + \ldots
+ \Delta E_n(\alpha) \,,                     \nonumber\\
\label{finres}
m_n^*(\alpha) &=& m_0^*(\alpha) + \Delta m^*_2(\alpha) + \ldots
+ \Delta m^*_n(\alpha) \,,                          \\
N_n(\alpha) &=& N_0(\alpha) + \Delta N_2(\alpha) + \ldots
+ \Delta N_n(\alpha) \,.                     \nonumber
\end{eqnarray}
In particular,
\begin{eqnarray}
\Delta E_2(\alpha) &=& - \frac{\alpha^2 \pi}{2\beta}
\int\!\!\!\!\int\!\!\!\!\int\limits_{0}^{\beta}\!\!\!\!\int\!\! dtds
dxdy \,\frac{e^{-|t-s|-|x-y|}}{\rho^{1/2}(t,s,x,y)} \,,  \nonumber\\
\Delta m^*_2(\alpha) &=& \frac{\alpha^2 \pi}{\beta}
\int\!\!\!\!\int\!\!\!\!\int\limits_{0}^{\beta}\!\!\!\!\int\!\!
dtdsdxdy \, \frac{e^{-|t-s|-|x-y|}}{\rho^{3/2}(t,s,x,y)}
\left\{(t-s)^2 F(x-y) \right.                           \nonumber\\
\label{secord}
&+& \left. (x-y)^2 F(t-s)+|t-s|\,|x-y|\,\Xi (t,s,x,y) \right\} \,, \\
\rho(t,s,x,y) &=& 4~F(t-s)F(x-y)-\Xi^2(t,s,x,y)\,,\nonumber\\
\Xi(t,s,x,y)  &=& F(t-x)+F(s-y)-F(s-x)-F(t-y)  \,.\nonumber
\end{eqnarray}
Note, the integrations over $s$ and $y$ in (\ref{secord}) may be taken
explicitly. The remaining double integrals can be derived analytically
in the weak- and strong-coupling limit, and numerically for finite
$\alpha$.

%%%%%%%%%%%%%%%%%%%%%%%%%%%%%%%%%%%%%%%%%%%%%%%%%%%%%%%%%%%%%%%%%%%%%
\section{Exact and Numerical Results}
%===================================================================

We admit that the accuracy reached with the second- and third-order
non-Gaussian corrections in (\ref{finres}) is sufficient for our
consideration. Below we have represented exact analytical solutions
in the weak- and strong-coupling limit and reasonable numerical data
for finite $\alpha$.

%%%%%%%%%%%%%%%%%%%%%%%%%%%%%%%%%%%%%%
\vskip 2mm
{\bf Weak coupling}
\vskip 2mm
%%====================================

 For $\alpha\to 0$ the solution of the constraint equations
(\ref{consteq}) reads
\begin{equation}
\widetilde{D}(k) = \left\{ k^2 +
2 \, \alpha \left( \sqrt{2} \, \sqrt{ 1 + \sqrt{1+k^2} } - 2 \right)
+ \alpha^2 \, \chi(k^2) \right\}^{-1} + O(\alpha^3) \,,
\label{weakasym1}
\end{equation}
where $\chi(k^2)$ is a monotonically increasing positive function.
Then, we obtain the exact GSE, EM and ANVP of the one-dimensional
polaron up to the $\alpha^3$-order. The results read
\begin{eqnarray}
&& E_3(\alpha) = - \, \alpha -\left( {3\over\sqrt{8}}-1 \right)
\,\alpha^2 - \left( 5-{63\sqrt{2} - 19\sqrt{3}\over 16}
- {29\sqrt{6}\over 48} \right) \,\alpha^3 - O(\alpha^4)\,, \nonumber\\
\label{e3m3weak}
&& m^{*}_3(\alpha) = 1 + {1\over 2}\,\alpha
+ \left( {5\over 8\sqrt{2}}-{1\over 4} \right) \, \alpha^2
+ 0.0691096281\,\alpha^3 + O(\alpha^4)  \,,                 \\
&& N_3(\alpha) = {1\over 2}\,\alpha +\left( {3\over\sqrt{2}}-2 \right)
\,\alpha^2 -{7\over 2}\left( 5-{63\sqrt{2} - 19\sqrt{3}\over 16}
- {29\sqrt{6}\over 48} \right) \,\alpha^3 + O(\alpha^4) \,.  \nonumber
\end{eqnarray}

 A refined weak-coupling method based on the three-phonon correction
has resulted in \cite{quin94}
\begin{eqnarray}
E_{CWW}(\alpha) &=& - \,\alpha - 0.06066\,\alpha^2
- 0.00844\,\alpha^3 \,,               \nonumber \\
\label{cww}
m^*_{CWW}(\alpha) &=& 1+0.5 \, \alpha + 0.19194\,\alpha^2
- 0.06912\,\alpha^3 \,,                 \\
N_{CWW}(\alpha) &=& 0.5 \,\alpha + 0.12132\,\alpha^2
+ 0.02954 \,\alpha^3 \,.               \nonumber
\end{eqnarray}

Our weak-coupling results (\ref{e3m3weak}) improve the known data
\cite{quin94,peet91} on the one-dimensional polaron. Note, a recent
attempt by considering an additional confining dot potential has
resulted in the binding energy and effective mass diverging in one
dimension \cite{saho98}.

%%%%%%%%%%%%%%%%%%%%%%%%%%%%%%%%%%%%%%
\vskip 2mm
{\bf Strong coupling}
\vskip 2mm
%%====================================

 In the strong-coupling limit the solution of (\ref{consteq}) looks
like
\begin{equation}
\widetilde{D}(k)=(k^2+\mu^2)^{-1} \,, \qquad \mu = {4\alpha^2/\pi} \,.
\label{strongasym1}
\end{equation}
The Gaussian leading-order and the corrected estimates for the GSE,
EM and ANVP become
\begin{eqnarray}
E_0(\alpha) &=&   -(1/\pi)  \,\alpha^2 \,,  \qquad ~~~~
m^{*}_0(\alpha) = (16/\pi^2) \,\alpha^4 \,,  \qquad ~~
N_0(\alpha)     =  (2/\pi) \,\alpha^2 \,,  \nonumber\\
\label{e3m3stro}
E_2(\alpha) &=& - 0.327014 \,\alpha^2 \,,  \qquad
m^{*}_2(\alpha) = 1.858065 \,\alpha^4 \,,  \qquad
N_2(\alpha)     = 0.654028 \,\alpha^2 \,,  \\
E_3(\alpha) &=& - 0.330205 \,\alpha^2 \,,  \qquad
m^{*}_3(\alpha) = 1.966430 \,\alpha^4 \,,  \qquad
N_3(\alpha)     = 0.660410 \,\alpha^2 \,.  \nonumber
\end{eqnarray}

 Rigorous studies of the strong-coupling regime of the polaron in one
dimension have been performed, particularly, in \cite{gros76,lesc89}.
Within the Pekar adiabatic approximation \cite{peet91} efficient for
$\alpha\to\infty$, one arrives at the following variational task
with a normalized wave function $\Psi(x)$:
\begin{eqnarray}
E_A(\alpha) = - \alpha^2 \!\!\int\! dx \left\{ \Psi(x)\Delta\Psi(x)
+ 2 \, |\Psi(x)|^4 \right\} \,, \qquad
m_A^{*}(\alpha) = \alpha^4 \,{4\over\pi} \!\!\int\! dk\,k^2\,
|\langle \Psi |e^{ikx}| \Psi \rangle|^2 \,.
\label{pekarener}
\end{eqnarray}
A modified polynomial placed instead of $\Psi(x)$ in \cite{peet91} has
resulted in
$E_{PS}=-0.3330877\,\alpha^2$, $m_{PS}^{*}=2.1254\,\alpha^4$ and
$N_{PS}=0.6661754\,\alpha^2$. Fortunately, problem (\ref{pekarener})
admits explicit analytic solutions (obtained first in \cite{gros76}
with a re-scaled coupling) as follows:
\begin{eqnarray}
\Psi(x) = \sqrt{2} / (e^{x} + e^{-x})   \,, \quad
E_A(\alpha) = - \alpha^2/3       \,, \quad
m_A^{*}(\alpha) = 32\,\alpha^4/15  \,, \quad
N_A(\alpha) = 2\,\alpha^2/3  \,.
\label{pekarexact}
\end{eqnarray}
 We see that our Gaussian (leading-order) approximations $E_0(\alpha)$,
$m^*_0(\alpha)$ and $N_0(\alpha)$ differ from the adiabatic results
(\ref{pekarexact}). However, these gaps have been systematically
narrowed by involving higher-order non-Gaussian contributions in
(\ref{e3m3stro}).

%%%%%%%%%%%%%%%%%%%%%%%%%%%%%%%%%%%%%%
\vskip 2mm
{\bf Intermediate coupling}
\vskip 2mm
%%====================================

 For finite $\alpha$ we have solved numerically the constraint
equations in (\ref{consteq}). The obtained numerical results for the
Gaussian leading-order contributions $E_0(\alpha)$ and $m_0^*(\alpha)$
as well as for the next-to-leading corrected data are plotted in Fig. 1
in comparison with the Feynman estimates \cite{dega86}.  In Fig. 2 we
have represented $N_0(\alpha)$ calculated by formula (\ref{number0}).
According to our results, the number of virtual phonons increases fast
as $\alpha$ grows and hence, the three-phonon correlation method
\cite{quin94} should be used carefully for $\alpha\ge 2.5$.

%%---------------------FIGURE 1----------------------------------------
\begin{figure}[ht]
\begin{center}
\epsfig{figure=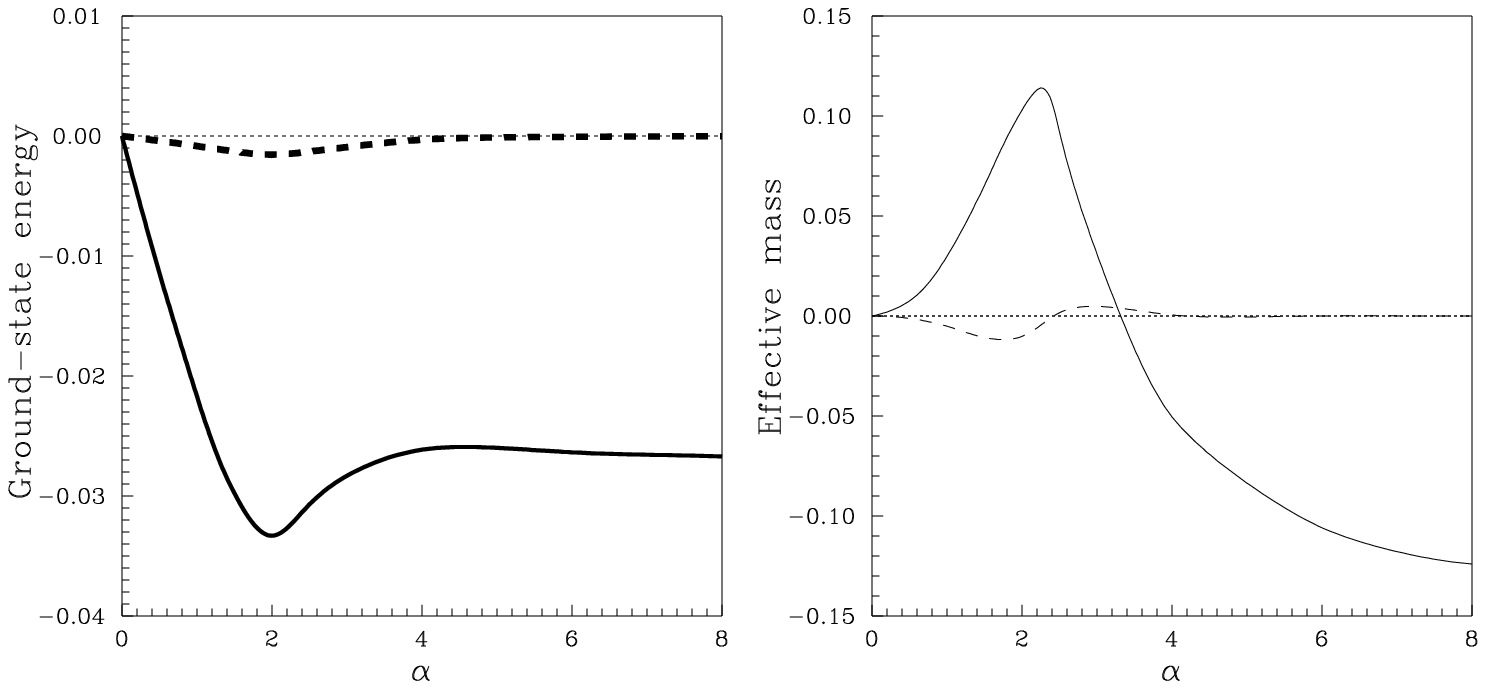,width=0.65\textwidth}

                         Fig. 1 \\[4mm]
\begin{minipage}{14cm}
{The ground-state energy and effective mass of the
one-dimensional polaron normalized to the corresponding Feynman
estimates (dotted horizontal lines) versus the coupling constant
$\alpha$. Dashed lines represent the leading-order Gaussian
approximations and solid curves correspond to the corrected results
taking into account the second-order non-Gaussian contributions.}
\end{minipage}
\end{center}
\end{figure}
%%---------------------------------------------------------------------

 Note, that the Feynman-type estimation \cite{dega86} represents a
partial case of the Gaussian leading-order approximation, when a trial
convex function $\widetilde{D}_F(k)= w/k^2+(1-w)/(k^2+\mu^2)$ is used
instead of the exact solution $\widetilde{D}(k)$. Therefore, the
Feynman estimate is less exact than the general Gaussian result, in
particular, $E_{0F}(\alpha)>E_{0}(\alpha)$.  \\[1mm]

%%---------------------FIGURE 2----------------------------------------
\begin{figure}[ht]
\begin{center}
\epsfig{figure=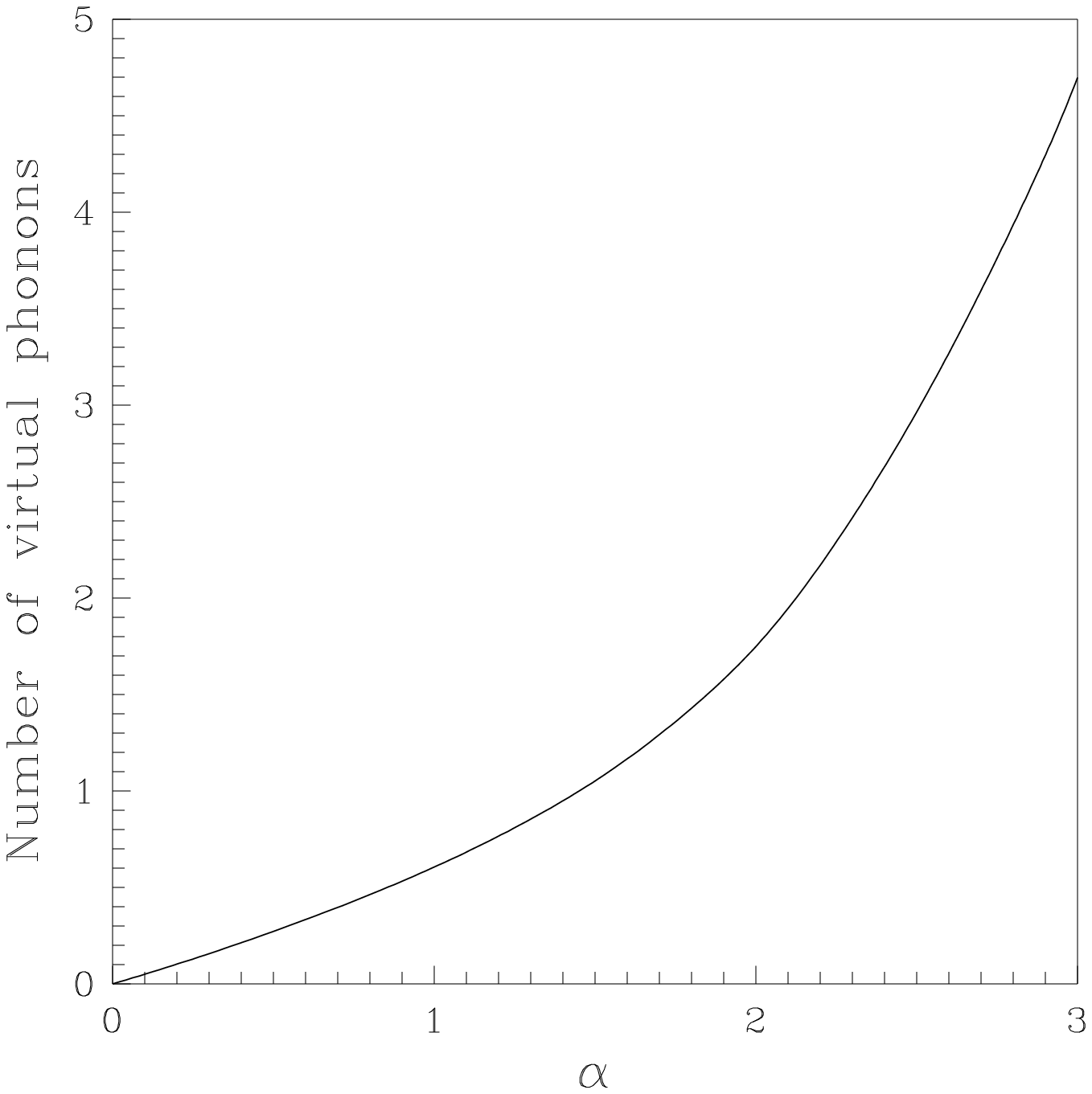,width=0.4\textwidth}

                         Fig. 2 \\[4mm]
\begin{minipage}{14cm}
{The Gaussian approximation to the average number of virtual
phonons $N_0(\alpha)$ versus the coupling constant $\alpha$.}
\end{minipage}
\end{center}
\end{figure}
%%---------------------------------------------------------------------

 In conclusion, the ground-state characteristics of the large optical
polaron confined in one dimension have been calculated by using the
generalized Gaussian approximation method. Exact analytic solutions
for the GSE, EM and ANVP have been obtained up to the order $\alpha^3$
in the weak-coupling limit. In the strong coupling limit, our Gaussian
leading-order results deviate (e.g., by 0.9 percent for the energy)
from the adiabatic data obtained within the Pekar Ansatz, but the gaps
have been systematically narrowed by involving the higher-order
corrections. For finite coupling, the leading-order Gaussian GSE
improves the Feynman variational estimate \cite{dega86} and represents
the lowest upper bound available for the one-dimensional polaron.
Our result for the number of excited phonons limits the validity
region of the few-phonon approximation methods. The next-to-leading
non-Gaussian contributions have been calculated to correct the
Gaussian approximations for finite $\alpha$.

Applied to the polaron confined in one dimension, the GGA method
provides reliable results rather quickly in the entire range of
coupling. Therefore, it will be reasonable to apply the GGA method
to improve the Lieb-Yamazaki type lower-bound estimate
(see, e.g. \cite{ganb00}) of the polaron self-energy.

%%========================================
\vskip 5mm
{\bf Acknowledgments \\[1mm]}
%%----------------------------------------

 The author is indebted to G.V.~Efimov and H.~Leschke for useful
discussions. Special thanks go to H.L. for helpful comments.

%-----------------------------------------------------------------

%-----------------------------------------------------------------
\end{document}